\documentstyle [12pt] {article}

\newcommand{\gtwid}{\mathrel{\raise.3ex\hbox{$>$\kern-.75em\lower1ex
\hbox{$\sim$}}}}
\newcommand{\ltwid}{\mathrel{\raise.3ex\hbox{$<$\kern-.75em\lower1ex
\hbox{$\sim$}}}}
\newcommand{\beq}{\begin{equation}}
\newcommand{\eeq}{\end{equation}}
\newcommand{\beqs}{\begin{eqnarray}}
\newcommand{\eeqs}{\end{eqnarray}}

\catcode`@=11
\@addtoreset{equation}{section}
\@addtoreset{equation}{subsection}
\def\theequation{\ifnum\value{section}=0 \arabic{equation}\ignorespaces
\else \ifnum\value{section}=-1 A.\arabic{equation}\ignorespaces 
\else \ifnum\value{subsection}=0 \thesection.\arabic{equation}\ignorespaces
\else \thesection.\arabic{subsection}.\arabic{equation}\ignorespaces
                           \fi
                      \fi
                 \fi}
\catcode`@=12

\begin{document}

\def\thefootnote{\fnsymbol{footnote}}
\baselineskip 6.0mm

\begin{flushright}
\begin{tabular}{l}
ITP-SB-94-54    \\
December, 1994  
\end{tabular}
\end{flushright}

\vspace{8mm}
\begin{center}
{\Large \bf Complex-Temperature Singularities in the $d=2$ Ising}
\vspace{3mm}
{\Large \bf Model: III. Honeycomb Lattice}
\vspace{4mm}
\vspace{16mm}

\setcounter{footnote}{0}
Victor Matveev\footnote{email: vmatveev@max.physics.sunysb.edu}
\setcounter{footnote}{6}
and Robert Shrock\footnote{email: shrock@max.physics.sunysb.edu}

\vspace{6mm}
Institute for Theoretical Physics  \\
State University of New York       \\
Stony Brook, N. Y. 11794-3840  \\

\vspace{20mm}

{\bf Abstract}
\end{center}

We study complex-temperature properties of the uniform and staggered 
susceptibilities $\chi$ and $\chi^{(a)}$ of the Ising model on the honeycomb
lattice.  From an analysis of low-temperature series expansions, we find 
evidence that $\chi$ and $\chi^{(a)}$ both have divergent singularities at the 
point $z=-1 \equiv z_{\ell}$ (where $z=e^{-2K}$), with exponents 
$\gamma_{\ell}'= \gamma_{\ell,a}'=5/2$. The critical amplitudes at this 
singularity are calculated. Using exact results, we extract the behaviour of 
the magnetisation $M$ and specific heat $C$ at complex-temperature
singularities. We find that, in addition to its zero at the physical critical 
point, $M$ diverges at $z=-1$ with exponent $\beta_{\ell}=-1/4$, vanishes 
continuously at $z=\pm i$ with exponent $\beta_s=3/8$, and vanishes 
discontinuously elsewhere along the boundary of the complex-temperature 
ferromagnetic phase.  $C$ diverges at $z=-1$ with exponent $\alpha_{\ell}'=2$
and at $v=\pm i/\sqrt{3}$ (where $v = \tanh K$) with exponent $\alpha_e=1$, 
and diverges logarithmically at $z=\pm i$. We find that the exponent relation 
$\alpha'+2\beta+\gamma'=2$ is violated at $z=-1$; the right-hand side is 4
rather than 2. The connections of these results with complex-temperature 
properties of the Ising model on the triangular lattice are discussed. 

\vspace{35mm}

\pagestyle{empty}
\newpage

\pagestyle{plain}
\pagenumbering{arabic}
\renewcommand{\thefootnote}{\arabic{footnote}}
\setcounter{footnote}{0}

\section{Introduction}
\label{intro}

   There are several reasons for studying the properties of statistical
mechanical models with the temperature variable generalised to take on complex
values.  First, one can understand more deeply the behaviour of various 
thermodynamic quantities by seeing how they behave as analytic functions of
complex temperature.  Second, one can see how the physical phases of a given 
model generalise to regions in appropriate complex-temperature variables. 
Third, a knowledge of the complex-temperature singularities of quantities 
which have not been calculated exactly helps in the search for exact, 
closed-form expressions for these quantities.  This applies, in particular, 
to the susceptibility of the 2D Ising model, which has, to this day, never
been calculated, in contrast to the (zero-field) free energy, first calculated
(for the square lattice) by Onsager \cite{ons}, and the spontaneous 
magnetisation, the expression for which (for the square lattice) was 
proposed by Onsager and first calculated by Yang \cite{yang}.  A fourth reason
for the interest in complex-temperature singularities is that they can 
significantly influence the behaviour of a given quantity for physical values
of the temperature.  Indeed, early studies \cite{g69,dg,g75} of such 
complex-temperature singularities were motivated in part by the fact that 
when they occurred closer to the origin, in a certain low-temperature 
expansion variable, than the physical critical point, they precluded the 
application of the ratio test, then in common use, to determine the location 
of this critical point.  The first work on natural boundaries of the free
energy of the Ising model (on the square lattice) for complex temperature 
was in Ref. \cite{fisher}. 

     Several years ago, we reported some results on complex-temperature
singularities of the susceptibility and correlation length for the Ising 
model \cite{ms}, including a discussion of complex-temperature symmetries and 
a proof (for the square lattice) that the (zero-field) susceptibility can 
have at most finite non-analyticities on the border of the 
complex-temperature extension of the symmetric phase, apart from its 
divergence at the physical critical point.  Two recent papers have reported 
results on complex-temperature singularities for the Ising model on the 
square lattice \cite{egj, chisq}.  In Ref. \cite{chitri}, the present authors
extended the study by Guttmann \cite{g75} of complex-temperature properties of
the Ising model on the triangular lattice.  In the present paper we shall 
carry out a study of the complex-temperature properties of the Ising model 
on the honeycomb ($=$ hexagonal) lattice.  For completeness we note that an
analysis of complex-temperature singularities in the 3D Ising model, after 
those of Refs. \cite{g69,dg}, was given in Ref. \cite{ipz}. 

\section{Complex-Temperature Extensions of Physical Phases}
\label{general}

   Our notation follows that in our previous papers \cite{chisq,chitri}, so we
review it here only briefly.  We consider the Ising model on the honeycomb
lattice (coordination number $q=3$) at a temperature $T$ and 
external magnetic field $H$ defined by the partition function 
$Z = \sum_{\{\sigma_i\}} e^{-\beta {\cal H}}$ with the Hamiltonian
\beq
{\cal H} = -J \sum_{<ij>} \sigma_i \sigma_j - H \sum_i \sigma_i
\label{ham}
\eeq
where $\sigma_i = \pm 1$ are the $Z_2$ spin variables on each site $i$ of the
lattice $\beta = (k_BT)^{-1}$, and $J$ is the exchange constant.  We use the
standard notation $K = \beta J$, $h = \beta H$, $v = \tanh K$, 
$z = e^{-2K} = (1-v)/(1+v)$, and $u=z^2$.  Another relevant variable is 
the elliptic modulus which one encounters in the broken-symmetry phases, 
\beq
k_< = \frac{4z^{3/2}(1-z+z^2)^{1/2}}{(1-z)^3(1+z)}
\label{kl}
\eeq
and its inverse, which occurs in expressions in the $Z_2$-symmetric phase, 
\beq
k_> = k_<^{-1} = \frac{4v^3}{(1-v^2)^{3/2}(1+3v^2)^{1/2}}
\label{kg}
\eeq
We note the symmetries
\beq
K \to -K \ \Rightarrow \ \{ v \to -v \ , \ \ z \to 1/z \ , \ \ u \to 1/u \ ,
\ \ k_x \to -k_x \}
\label{ksym}
\eeq
where $k_x = k_<$ or $k_>$.  The reduced free energy per site is 
$f = -\beta F = \lim_{N_s \to \infty} N_s^{-1} \ln Z$ in the thermodynamic
limit, where $N_s$ is the number of sites on the lattice. The 
zero-field susceptibility is 
$\chi =\frac{\partial M(H)}{\partial H}|_{H=0}$, where $M(H)$ denotes the 
magnetisation.  The staggered susceptibility is denoted $\chi^{(a)}$.  
It is convenient to deal with the reduced quantities $\bar\chi = 
\beta^{-1}\chi$ and $\bar\chi^{(a)} = \beta^{-1}\chi^{(a)}$. 
We recall that on a loose-packed lattice such as the
honeycomb lattice, in the symmetric, paramagnetic (PM) phase, the uniform and 
staggered susceptibilities are simply related according to 
\beq
\bar\chi^{(a)}(v) = \bar\chi(-v)
\label{chivrel}
\eeq
 Following Onsager's solution for $f(K,h=0)$ on the square lattice, the free 
energy was calculated for the honeycomb lattice in the papers of 
Ref. \cite{fhc}.  The spontaneous magnetisation for the honeycomb lattice was
first given by Naya \cite{naya}.  The critical coupling
separating the symmetric, paramagnetic (PM) high-temperature phase from the 
phase with spontaneously broken $Z_2$ symmetry and ferromagnetic (FM) 
long-range order is $K_c = (1/4)\ln 3$, so that 
$v_c = 1/\sqrt{3}=0.577350...$ and $z_c=2-\sqrt{3}=0.267949..$. 
As usual for loose-packed lattices, the critical point
separating the PM phase from the phase with antiferromagnetic (AFM) long-range
order is $K=-K_c$, or equivalently, $v=-v_c$, $z=1/z_c=2+\sqrt{3}$. 

   We begin by discussing the phase boundaries of the model as a function of
complex temperature, i.e. the locus of points across which the free energy is
non-analytic.  As noted in Ref. \cite{ms}, 
there is an infinite periodicity in complex $K$ under certain shifts along the
imaginary $K$ axis as a consequence of the fact that the spin-spin interaction
$\sigma_i\sigma_j$ in ${\cal H}$ is an integer.  In particular, there is an
infinite repetition of phases as functions of complex $K$; these repeated 
phases are reduced to a single set by using the variables $v$ or $z$ owing 
to the symmetry relation $K \to K+ n i \pi \Rightarrow \{ v \to v, \ \ 
z \to z \}$. 
The requisite complex extensions of the physical phases can be seen by
using the exact expression for the free energy \cite{fhc},
\beq
f = \ln 2 + \frac{1}{4}\int_{-\pi}^{\pi}\int_{-\pi}^{\pi} \frac{d\theta_1
d\theta_2}{(2\pi)^2} \ln \Bigl \{ \frac{1}{2}  \Bigl [\cosh^3(2K) + 1 - 
\sinh^2(2K)P(\theta_1,\theta_2) \Bigr ] \Bigr \}
\label{fhc}
\eeq
where
\beq
P(\theta_1,\theta_2) = \cos \theta_1 + \cos \theta_2 +
\cos(\theta_1 + \theta_2)
\label{ptheta}
\eeq
The boundaries of the complex-temperature phases are comprised of 
the locus of points where the 
argument of the logarithm in $f$ vanishes.\footnote{
The free energy is trivially infinite at $K=\infty$; since this is an isolated
point and hence does not form part of a boundary separating phases, it
will not be important here.} Expressed in terms of the low-temperature 
variable $z$, $f$ is given by 
\beq
f = \frac{q}{2}K + 
\frac{1}{4}\int_{-\pi}^{\pi}\int_{-\pi}^{\pi} \frac{d\theta_1
d\theta_2}{(2\pi)^2} \ln \Bigl [\bigl (1+z \bigr)^2 \Bigl \{ 
(1-2z+6z^2-2z^3+z^4) -2z(1-z)^2P(\theta_1,\theta_2) \Bigr \} \Bigr ]
\label{flow}
\eeq
(where $q=3$). Evidently, the argument of the logarithm vanishes along the 
curve defined by the solutions to the equation 
\beq
(1-2z+6z^2-2z^3+z^4) -2z(1-z)^2 x = 0 \ \ , \qquad -\frac{3}{2} \le x \le 3
\label{zeq}
\eeq
where $x=P(\theta_1,\theta_2)$. This curve is shown in Fig. 1(a). 
(Note that the curve contains the point $z=-1$ where the initial factor 
$(1+z)^2$ vanishes.) 
Since eq. (\ref{zeq}) has real coefficients, the solutions are either real 
or consist of complex conjugate pairs, which explains the reflection symmetry 
of the curve in Fig. 1(a) about the real axis in the $z$ plane.  Furthermore, 
under the transformation $z \to 1/z$, the left-hand side of eq. (\ref{zeq}) 
retains its form, up to an overall factor of $z^{-6}$; consequently the 
locus of solutions given by the curve in Fig. 1(a) is invariant under this 
mapping, $z \to 1/z$.  For $x=3$, eq. (\ref{zeq}) has double roots at 
$z=z_c, \ 1/z_c$.  As $x$ decreases from 3 to $-1$, the pairs of complex 
solutions move along the curve, rejoining again in two pairs of double roots 
at $z=\pm i$ for $x=-1$.  Finally, as $x$ decreases from $-1$ 
to $-3/2$, the solutions move outward from $z=\pm i$ along the
unit circle; the leftward-moving roots join at $z=-1$ while the 
rightward-moving roots terminate at the endpoints $z=e^{\pm i \pi/3}$. 
We shall denote the leftmost ($\ell$) of the real roots as $z_{\ell}=-1$.  
The points $z=\pm i$ are singular points of the curve in the technical 
terminology of algebraic geometry \cite{alg}; specifically, they are 
multiple points of index 2, where two arcs of the curve cross each other 
(with an angle of $\pi/2$).  We denote these as $z_{s\pm}$, respectively.  
The endpoints at $z=e^{\pm i\pi/3}$ are, of course,
also singular points of the curve in
the mathematical sense. The physical phases are: (i) FM, for $0 \le z \le z_c$;
(ii) PM, for $z_c < z \le 1$; and (iii) AFM, for $1 < z \le \infty$.  The
complex-temperature extensions of these are the regions marked in Fig. 1(a). 
Note that the complex-temperature phase boundaries in Fig. 1(a) are 
formally the same as those for the Ising model on the triangular lattice, 
in the $v$ plane (shown in Fig. 1(c) in Ref. \cite{chitri}). This is a 
consequence of the geometric duality of the honeycomb and triangular 
lattices.  Recall, however, that the actual phase structures, both for 
physical and complex temperature, are different; in particular,
the model on the triangular lattice has no AFM phase, and the outer phase in
the $v$ plane (denoted O in Ref. \cite{chitri}) has no overlap with any
physical phase.

   The transformation from $z$ to the elliptic modulus $k_<$ given in
eq. (\ref{kl}) maps the $z$ plane to the image shown in Fig. 1(b).  As noted in
(\ref{ksym}), $z$ and $1/z$ are mapped to $k_<$ and $-k_<$, respectively.  The
point $z_c$ is mapped to $k_<=1$.  The three points $1/z_c$ and $\pm i$ are all
mapped to $k_<=-1$.  The point $z=0$, as well as the points at complex infinity
(i.e., $z= \lim_{\rho \to \infty} \rho e^{i\theta}$ for arbitrary real
$\theta$) are mapped to $k_<=0$.  In addition, given the factorisation 
$(1-z+z^2)=(z-e^{i\pi/3})(z-e^{-i\pi/3})$, it follows that the finite endpoints
$z=e^{\pm i\pi/3}$ are mapped to $k_<=0$.  This latter property has the
consequence that this elliptic modulus variable is not a useful one in which to
re-express the low-temperature series expansion for $\bar\chi$, in contrast to
the case of the square lattice, where the analogous variable $k_<$ played a
very valuable role.  The transformation given by (\ref{kl}) maps the
complex-temperature FM and AFM phases formally onto the same phase, 
occupying the
interior of the unit circle in the $k_<$ plane; similarly, it maps the
complex-temperature PM phase onto the exterior of this circle.  The complex
conjugate arcs of the unit circle from $\arg(z)=\pm \pi/3$ to $\arg(z)=\pm
\pi/2$ are mapped onto the same line segment extending along the negative real
axis in the $(k_<)$ plane from 0 to $-1$, while the complex conjugate arcs of
the unit circle from  $\arg(z)=\pm \pi/2$ to $z=-1$ are mapped onto the
negative real axis from $k_<=-1$ to $k_< = -\infty$.  As usual with a singular
point, the image of the point $z=-1$ depends on the direction in which one
approaches it in the $z$ plane; for example, if one approaches this point along
the negative real $z$ axis (with a infinitesimal positive imaginary part so 
that with the branch cut for the $z^{3/2}$ factor being placed along the
negative real axis, $(-1)^{3/2}$ evaluates to $-i$), the image approaches 
$k_< = -i \infty$, and so forth for other directions. 

   One may work out the phase boundaries for the honeycomb lattice in the $v$
plane either by transforming the boundaries in the $z$ plane using the 
bilinear conformal mapping $v=(1-z)/(1+z)$, or directly by analysing the 
free energy expressed in terms of $v$, viz.,
\beq
f = \ln 2 -\frac{3}{4}\ln(1-v^2) + 
\frac{1}{4}\int_{-\pi}^{\pi}\int_{-\pi}^{\pi} \frac{d\theta_1
d\theta_2}{(2\pi)^2} \ln \Bigl [1+3v^4 - 2v^2(1-v^2)P(\theta_1,\theta_2) 
\Bigr ]
\label{fhigh}
\eeq
(The second term is just the usual $(q/2)\ln(\cosh K)$ term.)  Aside from the
trivial infinity at $v=1$ ($K=\infty$), the locus of singularities of $f$
is given by the solutions to the equation $1+3v^4-2v^2(1-v^2)x=0$ for $-3/2 \le
x \le 3$. This is shown in Fig. 1(c).  This equation has real coefficients and
is invariant under the transformation $v \to -v$; the respective consequences
of these two properties are that the solution curve in Fig. 1(c) is invariant
under reflection about the horizontal and vertical axes in the $v$ plane. 
The vertical line segments extend from $\pm i/\sqrt{3}$ to 
$\pm i \infty$, respectively, and are the images, under the conformal mapping,
of the portion of boundary in the $z$ plane lying on the unit circle. 
The curve in Fig. 1(c) is formally the same as the complex-temperature phase 
boundary for the Ising model on the triangular lattice in the variable $z$, 
because of the duality noted above. 

    Using the general fact that the high-temperature and 
(for discrete spin models such as
the Ising model) the low-temperature expansions have finite radii of
convergence, we can use standard analytic continuation arguments to establish
that in addition to the free energy and its derivatives, also 
the magnetisation and susceptibility
are analytic functions within each of the complex-temperature phases.  This
defines these functions as analytic functions of the respective complex
variable ($K$, $v$, or $z$, etc.).  Of course, 
these functions are, in general, complex away from the physical line 
$-\infty < K < \infty$.  

    Our definition of singular forms of a function at a complex-temperature 
singular point was given in Ref. \cite{chisq, chitri}.  Note, in particular,
that whereas a physical critical point can only be approached from two
different phases, high- or low-temperature, some complex-temperature 
singular points may be approached from more than two phases.  

    The low-temperature series for the staggered susceptibility is expressed in
terms of the variable $y=1/z$, and for our analysis of this series, we observe 
that the complex-temperature phase diagram in the $y$ plane has the same phase
boundaries as those in Fig. 1(a), owing to the invariance of this boundary
under $z \to 1/z$.  The phases are, of course, inverted, so that the innermost
phase is AFM, to its right, PM, and, in the outer region, FM. 

    Finally, because of the star-triangle relations connecting the Ising 
model on the triangular and honeycomb lattices, the following exact relations
hold \cite{fisherchi}: 
\beq
\chi_t(u) = \frac{1}{2}\Bigl [ \chi_{_{hc}}(z') + \chi^{(a)}_{hc}(z') \Bigr ]
\label{fisherlow}
\eeq
where
\beq
u = \frac{z'}{1-z'+z'^2}
\label{uzp}
\eeq
and 
\beq
\chi_t(w) = \frac{1}{2}\Bigl [ \chi_{_{hc}}(v) + \chi^{(a)}_{hc}(v) \Bigr ] = 
  \frac{1}{2}\Bigl [ \chi_{_{hc}}(v) + \chi_{_{hc}}(-v) \Bigr ]
\label{fisherhigh}
\eeq
where
\beq
v^2 = \frac{w}{1-w+w^2}
\label{vw}
\eeq

\section{Complex-Temperature Behaviour of the Specific Heat}

\subsection{General}
\label{generalc}

    From the exact expression (\ref{flow}) for the free energy $f$ 
we calculate the specific heat in the low-temperature phase as 
\begin{displaymath}
k_B^{-1}K^{-2}C = -\frac{8z^2}{(1-z^2)^2} + 
\frac{[3-12(z+z^7)+28(z^2+z^6)-20(z^3+z^5)+18z^4]}{\pi(1+z)(1-z)^5(1-z+z^2)}
K(k_<)
\end{displaymath}
\beq
 - \frac{3(1-z)(1+z)}{\pi(1-z+z^2)}E(k_<)
\label{clow}
\eeq
where $K(k)=\int_{0}^{\pi/2}d\theta [1-k^2\sin^2\theta]^{-1/2}$ and
$E(k)=\int_{0}^{\pi/2}d\theta [1-k^2\sin^2\theta]^{1/2}$ are the complete
elliptic integrals of the first and second kinds, respectively, which depend 
on the (square of the) elliptic modulus $k$, and $k_<$ was given in eq. 
(\ref{kl}).\footnote{In passing, we note that Houtappel's
expressions for the internal energy and specific heat, eqs. (108) and (109),
respectively, in Ref. \cite{fhc}, are incorrect if one uses
the integrals $\epsilon_1(\beta)$ and $\epsilon_2(\beta)$ as he defines them,
with the range of integration from $\phi=0$ to $2\pi$.  If, instead, one takes
the range of integration from $\phi=0$ to $\phi=\pi/2$, so that the integrals
are just the usual elliptic integrals $K(\sqrt{\beta})$ and $E(\sqrt{\beta})$,
then his eqs. (108) and (109) become correct.}  The expression (\ref{clow})
applies in both the physical FM and AFM phases, and may be analytically
continued throughout the respective complex-temperature extensions of these
phases. 

   In the physical PM phase, a similar calculation starting from (\ref{fhc}) or
(\ref{fhigh}) yields 
\beq
k_B^{-1}K^{-2}C = v^{-2}(1-v^2)^{1/2} \biggl [ 
-\frac{1}{2}(1-v^2)^{3/2} + \frac{4}{\pi(1+3v^2)^{1/2}}K(k_>) 
-\frac{3(1-v^2)}{\pi(1+3v^2)^{1/2}}E(k_>) \biggr ]
\label{chigh}
\eeq
where $v=\tanh K$, and $k_>$ was given in (\ref{kg}). 
Having these exact expressions for the specific heat, we proceed to
work out its behaviour at complex-temperature singularities.  

\subsection{Vicinity of $z=-1$}

   As one approaches the point $z=-1$ (denoted $z_\ell$ as above) 
from within either the
complex-temperature FM or AFM phase, the specific heat diverges, 
with the dominant divergence arising from the first term in (\ref{clow}), which
becomes $-2(1+z)^{-2}$.  (There is also a weaker, logarithmic divergence from
the term involving $K(k_<)$.)  Hence, we find 
\beq
\alpha_{_{\ell,FM}}' = \alpha_{_{\ell,AFM}}' = 2
\label{alphaellp}
\eeq
Now, $K=-(1/2)\ln z$, so choosing the branch cut for the complex logarithm to
lie along the negative real axis and choosing the first Riemann sheet for the
evaluation of the logarithm, as $z$ approaches $-1$ from above or below the
negative real axis, one has $K_\ell = \mp i \pi/2$ respectively, and hence 
in both cases 
\beq
k_B^{-1}C \to \frac{\pi^2}{2(1+z)^2} \ \ , \qquad as \ \ z \to -1
\label{cm1}
\eeq
It is interesting to relate the critical exponent (\ref{alphaellp}) to the
critical exponent $\alpha_e'$ for the specific heat on the triangular 
lattice at the point $u=u_e=-1/3$, which corresponds, via (\ref{uzp}), to 
$z'=z=-1$ on the honeycomb lattice.  (Recall that the point $u=-1/3$ is an
endpoint of a singular line segment protruding into the complex-temperature FM
phase; hence, it can only be approached from within this phase, and so
$\alpha_e' \equiv \alpha_{e,FM}'$.)  In Ref. \cite{chitri}, using the exact
expression for the free energy on the triangular lattice, we calculated that
$\alpha_e'=1$.  Given the star-triangle relations which connect the Ising model
on these two lattices and the fact that the Taylor series expansion of 
$u+1/3$, as a function of $z'$, in the vicinity of $z'=-1$ ($=z$ on the
honeycomb lattice), starts with the quadratic term, 
\beq
u + \frac{1}{3} = \frac{1}{9}(1+z')^2 + \frac{1}{9}(1+z')^3 +O((1+z')^4)
\label{uzpm1}
\eeq
it follows that the exponents $\alpha_{_{\ell,FM}}'=\alpha_{_{\ell,AFM}}'=2$ 
at $z=-1$ on the honeycomb lattice have twice the value of $\alpha_{e}'=1$ 
at $u=-1/3$ on the triangular lattice.  
Note, however, that the leading divergence in $C$ arises
from different terms on the two different lattices (in the triangular lattice,
the leading divergence arises from the term proportional to $E(k_<)$, as 
discussed in Ref. \cite{chitri}). 

\subsection{Vicinity of $z = \pm i$}

   The points $z=\pm i$ can be approached from within the complex-temperature
extensions of the FM, AFM, and PM phases.  For the approach to $z=\pm i$ 
from within the complex FM and AFM phases, we find from (\ref{clow}) that the 
first term and the term involving $E(k_<)$ yield finite contributions, 
while the term involving $K(k_<)$ diverges logarithmically, 
as $\pm(4i/\pi)K(k_< \to -1)$.  Using the fact 
that as $\lambda \to \pm 1$, $K(\lambda) \to (1/2)\ln(16/(1-\lambda^2))$, and
the Taylor series expansion of $k_<^2$ in the neighborhood of $z=\pm i$, 
\beq
k_<^2 = 1 - 2(z \mp i)^3 + O((z \mp i)^4) 
\label{klsqi}
\eeq
we can express the most singular term on the RHS of eq. (\ref{clow}) as 
$\mp(2i/\pi)\ln[(z \mp i)^3]$.  Evaluating $K=-(1/2)\ln z$ for $z=\pm i$ on the
first Riemann sheet of the logarithm, we have $K = \mp i\pi/4$, so that 
\beq
k_B^{-1}C \sim \pm \frac{i}{8\pi}\ln \Bigl [ (z \mp i)^3 \Bigr ]
\label{ci}
\eeq
It follows that for $z=z_{s,\pm}=\pm i$, 
\beq
\alpha_{s,FM}'=\alpha_{s,AFM}'=0 \qquad (log. \ div.)
\label{alphasfmafm}
\eeq
These results are the same as we found for the specific heat exponent on the
triangular lattice at the point $u=-1$ corresponding, via (\ref{uzp}) with $z'
\equiv z$, to $z= \pm i$ on the honeycomb lattice. 

   For the approach to the points $v=v_{s\pm}=\pm i$ from within the 
complex-temperature PM phase, we use the expression for $C$ in this phase,
eq. (\ref{chigh}).  We find that the term involving $K(k_>)$ produces a 
logarithmically divergence in $C$, so that the specific heat exponent 
$\alpha_{s,PM} \equiv \alpha_s$ is 
\beq
\alpha_s = 0 \qquad (log. \ div.)
\label{alphaspm}
\eeq
Taking the branch cuts for the factor $(1+3v^2)^{1/2}$ to
lie along the semi-infinite line segments from $\pm i/\sqrt{3}$ to $\pm
i\infty$, and taking the approach such that $(-1)^{1/2}$ is evaluated as $+i$,
we find that this term yields $(2i/\pi)\ln[(1-k_>^2)]$.  Using the Taylor
series expansion of $k_>^2$, as a function of $v$, near $v=i$, 
\beq
k_>^2 = 1 -2i(v-i)^3 +O((v-i)^4)
\label{kgtaylor}
\eeq
and its complex conjugate for $v \to -i$, and the result 
$K={\rm arctanh}(\pm i) = \pm i\pi/4$, we find 
\beq
k_{B}^{-1}C \sim -\frac{i\pi}{8}\ln[(v \mp i)^3]
\label{cipm}
\eeq
(In the evaluation of the function ${\rm arctanh}(\zeta) = 
(1/2)\ln[(1+\zeta)/(1-\zeta)]$ here and below, we again use the first 
Riemann sheet of the logarithm.)

\subsection{Vicinity of $v = \pm i(3)^{-1/2}$}

   We next determine the singularities of the specific heat as one approaches
the endpoints $v=\pm v_e=\pm i/\sqrt{3}$ of the semi-infinite line
segments protruding into the complex-temperature PM phase.  Note that all
directions of approach to $v=\pm i/\sqrt{3}$ except exactly (down or up,
respectively) along the
singular line segments occur from within the complex PM phase. From 
(\ref{chigh}) we find that $C$ is
divergent, with the leading divergence arising from the term involving
$E(k_>)$.  This term gives $\pm (4\sqrt{3}/\pi)(1+3v^2)^{-1}$ as 
$v \to \pm i$, so 
\beq
\alpha_e = 1 
\label{alphae}
\eeq
Using $K={\rm arctanh}(\pm i/\sqrt{3}) = \pm i\pi/6$, we have
\beq
k_B^{-1}C \to \mp \frac{\pi}{3^{3/2}(1+3v^2)} \qquad as \ \ v \to 
\pm \frac{i}{\sqrt{3}} 
\label{cval}
\eeq

\subsection{Elsewhere on the Complex-Temperature Phase Boundary}

   The free energy $f(K)$ is non-analytic across the complex-temperature phase
boundaries, and hence, of course, this is also true of its derivatives with
respect to $K$, in
particular the internal energy $U$ and the specific heat $C$. As an
illustration, consider moving along a ray outward from the origin of the
$z$ plane defined by $z=r e^{i\theta}$ with $\theta < \pi/2$.  For a given
$\theta$, as $r$ exceeds the critical value $r_c(\theta)$, one passes from the
complex-temperature FM phase into the complex-temperature PM phase.  At the
phase boundary, as is clear from Fig. 1(b), the elliptic modulus $k_<$ has 
magnitude unity and can be written $k_< = e^{i\phi}$, where the angle $\phi$
depends on $\theta$. As we discussed in connection with Fig. 1(b), $z=z_c$ is
mapped to $k_<=1$, and $z=i$ to $k_<=-1$; $\phi$ increases from 0 at $\theta=0$
to $\pi$ at $\theta=\pi/2$.  Hence, for $0 < \theta < \pi/2$, $k_<$ has a
nonzero imaginary part.  Now when one passes through the FM-PM phase boundary
along the ray at this angle $\theta$, one changes the argument of the elliptic
integrals from $k_<=e^{i\phi}$ to $k_>=1/k_<=e^{-i\phi}$.  The elliptic
integrals $K(k)$ and $E(k)$ are analytic functions of $k^2$ with, respectively,
a logarithmically divergent and a finite branch point singularity 
at $k^2=1$ and associated
branch cuts which may be taken to lie along the positive real axis in the $k^2$
plane.  In particular, $K(k)$ and $E(k)$ are both analytic at the point 
$k=k_<=e^{i\phi}$ for $0 < \theta < \pi/2$.  Hence, when we replace the
argument $k_<$ by $k_>$, which is the complex conjugate of $k_<$ on the unit
circle, we have $F(k_>=e^{-i\phi}) = F(k_<=e^{i\phi})^*$ for $F=K,E$.  Since
these elliptic integrals are complex for generic complex $k_<$, it follows that
their imaginary part is discontinuous across the FM-PM boundary.  The
coefficients of the elliptic integrals are also different functions in the FM
and PM phases, and these coefficients are discontinuous as one
crosses the boundary between these phases on the above ray.  Combining these,
we find that the specific heat itself is discontinuous as one moves across the
FM-PM boundary on this ray. 

\section{Complex-Temperature Behaviour of the Spontaneous Magnetisation}

     The spontaneous magnetisation is given in the physical FM phase by 
\cite{naya}
\beqs 
M_{hc} & = & \bigl (1-(k_<)^2 \bigr )^{1/8} \nonumber \\
       &   & \nonumber \\
  & = & \frac{(1+z^2)^{3/8}(1-4z+z^2)^{1/8}}{(1-z)^{3/4}(1+z)^{1/4}} 
\label{mhc}
\eeqs
and vanishes identically elsewhere. Observe that 
$(1-4z+z^2)=(1-z/z_c)(1-z_cz)$. The expression 
(\ref{mhc}) for $M$ can be analytically continued throughout the 
complex-temperature extension of the physical FM phase.  It evidently 
vanishes continuously at both the physical critical point $z=z_c$, with
critical exponent $\beta=1/8$, and at the complex-temperature points 
$z = z_{s\pm} = \pm i$, with exponent 
\beq
\beta_s=\frac{3}{8}
\label{betas}
\eeq
As we have observed earlier \cite{chisq}, this exponent is the same as
the exponent $\beta_{s,t}=3/8$ 
characterising the zero in the magnetisation on the triangular
lattice at the point $u=-1$ corresponding, via eq. (\ref{uzp}), to $z'=z=\pm i$
on the honeycomb lattice; however, $\beta_{s,hc}=\beta_{s,t}$ differ from
$\beta_{s,sq}=1/4$ characterising the zero in $M$ on the square lattice at 
$u=-1$.  This is a violation of universality, since this critical exponent 
is evidently lattice-dependent.  A way of understanding the origin of this
violation was discussed in Ref. \cite{chisq} and directly reflects the fact
that the Hamiltonian and hence the internal energy are not, in general, real
numbers at complex-temperature singular points. 

  Furthermore, $M$ has a divergence at $z=z_{\ell}=-1$, with exponent 
\beq
\beta_{\ell} = -\frac{1}{4}
\label{betal}
\eeq
Note that the apparent zero at $z=1/z_c$ and the apparent divergence at $z=1$
do not actually occur, since these points are outside of the
complex-temperature extensions of the FM phase, in which the (analytic
continuation of the) formula (\ref{mhc}) applies. We recall that, as a 
consequence of the star-triangle relation which connects the Ising model on the
triangular and honeycomb lattices \cite{domb1}, 
\beq
M_{t}(u) = M_{hc}(z')
\label{magrel}
\eeq
where \cite{potts}
\beq
M_{t} = \biggl (\frac{1+u}{1-u}\biggr )^{3/8}
\biggl(\frac{1-3u}{1+3u}\biggr)^{1/8}
\label{mtri}
\eeq
and $z'$ ($=z$ on the honeycomb lattice) was given in eq. (\ref{uzp}). 

 In Table 2 we list a comparison of singularities in $M_t$ and $M_{hc}$. Note
that, aside from the points listed in this table, these functions vanish 
discontinuously along the borders of the respective complex-temperature FM 
phases.  The fact that the exponent $\beta_{\ell}$ with which $M_{hc}$ 
diverges at the point $z=-1$ on the honeycomb lattice is twice the 
exponent with which $M_t$ diverges at the point $u=-1/3$ on the triangular 
lattice (which corresponds to $z'=z=-1$ via (\ref{uzp})) follows from the 
star-triangle relation connecting the Ising model on these two lattices and the
property that the Taylor series expansion of $u+1/3$, as a function of 
$z'$, in the vicinity of $z'=-1$, starts with the quadratic term, as given in
eq. (\ref{uzpm1}). 

\begin{table}
\begin{center}
\begin{tabular}{|c|c|c|c|c|} \hline \hline  & & & & \\
$u$, \ $t$ & $z$, \ $hc$ & $M$ & $\beta$, \ $t$ & $\beta$, \ $hc$ \\
 & & & & \\
\hline \hline
$u_c=1/3$ & $z_c=2-\sqrt{3}$ & $0$ \ (cont.) & $1/8$ & $1/8$ \\ \hline
$-1$ & ${\pm i}$ & $0$ \ (cont.) & $3/8$ & $3/8$  \\ \hline
$-1/3$ & $-1$ & div. & $-1/8$ & $-1/4$ \\ \hline\hline
\end{tabular}
\end{center}
\caption{Comparative singularities of $M$ on the triangular ($t$) and 
honeycomb ($hc$) lattices.  Points are related according to the 
transformation (\ref{uzp}) with $z' = z$ ($hc$).  The notations $0$ 
\ (cont.) and div. denote, respectively, a point where $M$ vanishes
continuously and where it diverges.}
\end{table}

   Since the honeycomb lattice is loose-packed, one immediately infers the 
staggered magnetisation $M_{hc,st}$ from the (uniform) magnetisation $M_{hc}$:
formally, 
\beq
M_{hc,st}(y) = M_{hc}(z \to y)
\label{mstag}
\eeq
Hence, $M_{hc,st}$ has continuous zeros at both the physical critical point 
$y=y_c$ and the complex-temperature point $y=-1$, and diverges 
at the points $y=\pm i$. 

\section{Analysis of the Low-Temperature Series for $\bar\chi$ }
\label{singfm}

\subsection{General}
\label{generalchilow}

   Here we shall study the complex-temperature singularities of the
susceptibility $\bar\chi$ which occur as one approaches the boundary of 
the (complex-temperature extension of the) FM phase from within this phase.  
In the next section we shall carry out a similar analysis for the 
staggered susceptibility, $\bar\chi^{(a)}$.  In particular, we consider 
the behaviour in the vicinity of the points $z=-1$ and $z=\pm i$.  
For the study of $\bar\chi$, we use the low-temperature series expansion 
for $\bar\chi$, which is given by 
\beq
\bar\chi = 4z^3 \Bigl ( 1 + \sum_{n=1}^{\infty} c_n z^n \Bigr )
\label{chiseries}
\eeq
For the study of $\bar\chi^{(a)}$, we
use the corresponding low-temperature series expansion 
\beq
\bar\chi^{(a)} = 4y^3 \Bigl ( 1 + \sum_{n=1}^{\infty} c_{n,a} y^n \Bigr )
\label{chiaseries}
\eeq
where $y=1/z$ is the expansion variable in the AFM phase.  
These expansions have finite radii of convergence and, by analytic 
continuation from the physical low-temperature intervals $0 \le z < z_c$ and $0
\le y \le y_c$, apply throughout the complex-temperature extensions of the FM
and AFM phases, respectively.  (Here $y_c$ is the critical point separating the
PM and AFM phases.  As can be seen from Fig. 1(a), this occurs at
$z=1/z_c=2+\sqrt{3}$, so that $y_c=2-\sqrt{3}$, the same numerical value as the
critical point $z_c$ separating the PM and FM phases.)  Since the $z^3$ and
$y^3$ factors are known exactly in the respective series expansions
(\ref{chiseries}) and (\ref{chiaseries}), it is convenient to study the 
remaining factors $\bar\chi_r = \bar\chi/(4z^3)$ and 
$\bar\chi^{(a)}_r = \bar\chi^{(a)}/(4y^3)$.  For the honeycomb lattice,
the expansion coefficients $c_n$ and $c_{n,a}$ were calculated by the
King's College group to order $n=13$ (i.e., $\bar\chi$ and $\bar\chi^{(a)}$ 
to $O(z^{16})$) in 1971 \cite{tlow1}, and to order $n=18$ (i.e., $\bar\chi$ 
and $\bar\chi^{(a)}$ to $O(z^{21})$) in 1975 \cite{tlow2}.  We have checked 
and found that apparently these series have not been calculated to higher 
order subsequently \cite{gaunt,guttmann}.  We have analysed these series 
using dlog Pad\'{e} and differential approximants (a recent review of the
methods is given in Ref. \cite{tonyg}).  For the approach to 
$z=-1$ (denoted $z_{\ell}$) from within the complex-temperature extension of
the FM phase, we write the leading singularity of $\bar\chi$ as 
\beq
\bar\chi(z) \sim A_{\ell}'|1-z/z_{\ell}|^{-\gamma_{\ell}'}
( 1+a_{1,\ell}|1-z/z_{\ell}| + ... )
\label{singform}
\eeq
where $A_{\ell}'$ and $\gamma_{\ell}'$ denote, respectively, the critical 
amplitude and critical exponent, and the dots $...$ represent analytic 
confluent corrections.  Similarly, for the approach to this point, which in
this context we denote $y_{\ell}=1/z_{\ell}=-1$, from within the
complex-temperature extension of the AFM phase, we write the leading
singularity of $\bar\chi^{(a)}$ as 
\beq
\bar\chi^{(a)}(y) \sim A_{\ell,a}'|1-y/y_{\ell}|^{-\gamma_{\ell,a}'}
( 1+a_{1,\ell,a}|1-y/y_{\ell}| + ... )
\label{singforma}
\eeq
(As is clear from Fig. 1(a), this point $z=1/y=-1$ can only be approached 
from within the complex-temperature FM or AFM phases.) 
As in our earlier studies of complex-temperature
singularities of $\bar\chi$ in the Ising model on the square and triangular
lattices \cite{chisq,chitri}, we have not included non-analytic confluent
corrections to the scaling form in eq. (\ref{singform}) since, as we have
discussed before, previous work has indicated that they are very weak or 
absent for the usual critical point of the 2D Ising model. 

   Before proceeding, we consider the implications of the exact relation
(\ref{fisherlow}). Given that $\chi_t(u)$ has a singularity at $u=u_e=-1/3$, 
it follows that the sum $\chi_{_{hc}}(z')+\chi_{_{hc}}^{(a)}(z')$ on the 
honeycomb lattice has the same singularity at the point $z'=1$ 
corresponding via (\ref{uzp}) to $u=-1/3$.  But this does not, by itself, 
determine the singularities in the individual functions $\chi_{_{hc}}$ and
$\chi_{_{hc}}^{(a)}$ at this point.  If one could prove that both 
$\chi_{_{hc}}$ and $\chi_{_{hc}}^{(a)}$ necessarily have the same singularity
at $z'=-1$, then the relation (\ref{fisherlow}), together with the result that
$\chi_t(u) \sim (1+3u)^{-\gamma_e'}$ with $\gamma_e'=5/4$
\cite{g75,chitri}, and
the Taylor series expansion of $u+1/3$ as a function of $z'$ 
(i.e., $z$ on the honeycomb lattice)  in the vicinity of $z'=-1$,
eq. (\ref{uzpm1}), would imply that 
$\gamma_{\ell}'=\gamma_{\ell,a}'=2\gamma_e'=5/2$. 
However, although it is plausible that $\chi_{_{hc}}$ and 
$\chi_{_{hc}}^{(a)}$ do have the same singularities at $z=-1$, there is no 
simple relationship between the respective low-temperature series for these 
two functions, as is clear from the first few terms \cite{tlow1}, 
\beq
\bar\chi = 4z^3\Bigl [1+6z+27z^2+122z^3+516z^4+2148z^5+...\Bigr ]
\label{chiterms}
\eeq
and 
\beq
\bar\chi^{(a)} = 4y^3\Bigl [1+0 \cdot y+3y^2+2y^3+12y^4+24y^5+...\Bigr ]
\label{chiaterms}
\eeq
Hence, an explicit series analysis is worthwhile to obtain the critical 
exponents. 

\subsection{Exponent at $z=-1$ Singularity}

   Since, as expected, we obtained more precise results from the 
differential approximants than the dlog Pad\'{e} approximants, we shall 
concentrate on the former here.  Our notation follows that 
of Guttmann \cite{tonyg} and our earlier papers \cite{chisq,chitri}, so we 
shall only describe it briefly.  In this method, the function 
$f = \bar\chi_r(\zeta)$ being approximated satisfies a linear ordinary
differential equation (ODE) of $K'th$ order, ${\cal L}_{{\bf M},L}f_K(\zeta) =
\sum_{j=0}^{K}Q_{j}(\zeta)D^j f_K(\zeta) = R(\zeta)$, where
$Q_j(\zeta) = \sum_{\ell=0}^{M_j} Q_{j,\ell}\zeta^{\ell}$ and
$R(\zeta)=\sum_{\ell=0}^L R_\ell \zeta^\ell$ (and $\zeta$ denotes a generic
complex variable).  We shall use the implementation of the
method in which $D \equiv \zeta d/d\zeta$.  The solution to
this ODE, with the initial condition $f(0)=1$, is
the resultant approximant, labelled as $[L/M_0;...;M_K]$.  The general solution
of the ODE has the form $f_j(\zeta) \sim A_j(\zeta)|\zeta-\zeta_j|^{-p_j} +
B(\zeta)$ for $\zeta \to \zeta_j$.  The singular points $\zeta_j$ are
determined as the zeroes of $Q_K(\zeta)$ and are regular singular points of
the ODE, and the exponents are given by $-p_j = K-1-Q_{K-1}(\zeta_j)/(\zeta_j
Q_K'(\zeta_j))$.  We recall that if the series for the function is 
calculated to order $\zeta^N$, then one can compute the differential 
approximants $[L/M_0;M_1]$ up to order $L+M_0+M_1=N-2$.  We shall list the
results from the differential approximants from to $\bar\chi_r$ and
$\bar\chi^{(a)}_r$ from $L+M_0+M_1=13$ to the maximum value possible using the
series for these functions calculated to $O(z^{18})$ and $O(y^{18})$, viz.,
$L+M_0+M_1=16$.  We take $K=1$, which will be adequate for our purposes, and
use unbiased differential approximants since this allows us to apply an
extrapolation technique as in our earlier work.  In this technique, we plot the
value of the exponent obtained from each differential approximant as a function
of the distance of the corresponding pole location from the inferred exact 
position of the singularity (e.g., $z=-1$, etc.).  We then extrapolate to 
zero distance of the pole from this singularity to obtain the estimate of 
the exponent.  Of course, this is essentially equivalent
to using biased differential approximants.  We present our results in 
Table 2.

\begin{table}
\begin{center}

\begin{tabular}{|c|c|c|c|} \hline \hline  & & & \\
$[L/M_0;M_1]$ & $z_{sing}$ & $|z_{sing}-z_{\ell}|$ & $\gamma_{\ell}'$ \\
 & & & \\
\hline \hline
$[1/6;6]$ & $-1.0096461$ & $9.6 \times 10^{-3} $ & $ 2.5188$ \\ \hline
$[1/6;7]$ & $-1.0009020$ & $9.0 \times 10^{-5} $ & $ 2.3989$ \\ \hline
$[1/7;5]$ & $-1.0056732$ & $5.7 \times 10^{-3} $ & $ 2.4418$ \\ \hline
$[2/6;6]$ & $-1.0047536$ & $4.8 \times 10^{-3} $ & $ 2.4588$ \\ \hline
$[2/7;5]$ & $-0.9945961$ & $5.4 \times 10^{-3} $ & $ 2.2967$ \\ \hline
$[3/6;7]$ & $-0.9959243$ & $4.1 \times 10^{-3} $ & $ 2.4136$ \\ \hline
$[3/7;6]$ & $-0.9903576$ & $9.6 \times 10^{-3} $ & $ 2.3022$ \\ \hline
$[4/4;6]$ & $-0.9948731$ & $5.1 \times 10^{-3} $ & $ 2.1301$ \\ \hline
$[4/6;6]$ & $-0.9946423$ & $5.4 \times 10^{-3} $ & $ 2.3580$ \\ \hline
$[5/4;5]$ & $-1.0007133$ & $7.1 \times 10^{-5} $ & $ 2.1252$ \\ \hline
$[5/4;6]$ & $-1.0064564$ & $6.5 \times 10^{-3} $ & $ 2.0704$ \\ \hline

\end{tabular}
\end{center}
\caption{Values of $z_{sing}$ and $\gamma_{\ell}'$ from differential
approximants to low-temperature series for $\bar\chi_r(z)$. See text for
definition of $[L/M_0,M_1]$ approximant.  We only display entries which
satisfy the accuracy criterion $|z_{sing}-z_{\ell}| \le 10^{-2}$, where 
$z_{\ell}=-1$.}
\label{chitable}
\end{table}
These results yield evidence that $\bar\chi$ has a divergent singularity at
$z=-1$, as one approaches this point from the complex-temperature FM phase. 
Since the values of the exponent from the differential approximants show
considerable scatter, it is only possible to extract a rather crude 
estimate for $\gamma_{\ell}'$.  We obtain 
\beq
\gamma_{\ell}' = 2.4 \pm 0.2
\label{gammaell}
\eeq
This is consistent with the following inference which we shall make for the
exact value of this exponent:
\beq
\gamma_{\ell}'=\frac{5}{2}
\label{gammaellexact}
\eeq
We note that the low-temperature series does not yield as precise a
determination of this exponent at $z=-1$ on the honeycomb lattice as the 
corresponding low-temperature series did for the susceptibility exponent on the
triangular lattice at the point $u=-1/3$ related to $z=z'=-1$ via (\ref{uzp})
\cite{g75,chitri}. 
The critical amplitude $A_{\ell}'$ for the susceptibility at $z=-1$ will be 
discussed below. 

\section{Analysis of the Low-Temperature Series for $\bar\chi^{(a)}$}
\label{chia}

\subsection{General} 
\label{generalchia}

The staggered susceptibility $\bar\chi^{(a)}$ has a well-known divergent
singularity at $y=y_c=2$ with low-temperature exponent $\gamma^{(a)'}=7/4$. 
Here we analyse the complex-temperature singularities of 
this function using the low-temperature series given above in
eq. (\ref{chiaseries}). 

\subsection{Critical Exponent of $\bar\chi^{(a)}$ at $z=-1$ Singularity}
\label{gamma_a}

Our results from the differential approximants are listed in Table 3.  From
these we find strong evidence that as one approaches the point $y=1/z=-1$ 
from within the complex AFM phase, $\bar\chi^{(a)}$ has a divergent 
singularity.  It is interesting that the exponent values from these 
differential approximants 
show less scatter than those which we found for the uniform
susceptibility. Furthermore, there is a better correlation between the value of
the exponent and the distance of the pole location from the inferred exact 
value of the singularity in Table 3, as compared with Table 2.  Using our 
extrapolation technique, we obtain 
\beq
\gamma_{\ell,a}' = 2.50 \pm 0.02
\label{gammaella}
\eeq
This is consistent with the following exact value, which we infer:
\beq
\gamma_{\ell,a}' = \frac{5}{2}
\label{gammaellaexact}
\eeq
so that, with this inference, 
\beq
\gamma_{\ell,a}' = \gamma_{\ell}'
\label{gammaequal}
\eeq

\begin{table}
\begin{center}

\begin{tabular}{|c|c|c|c|} \hline \hline  & & & \\
$[L/M_0;M_1]$ & $y_{sing}$ & $|y_{sing}-y_{\ell}'|$ & $\gamma_{\ell,a}'$ \\
 & & & \\
\hline \hline
$[0/7;6]$ & $-0.9987578$ & $1.2 \times 10^{-3} $ & $ 2.4468$ \\ \hline
$[0/7;7]$ & $-0.9901632$ & $9.8 \times 10^{-3} $ & $ 2.3222$ \\ \hline
$[0/7;8]$ & $-0.9993457$ & $6.5 \times 10^{-4} $ & $ 2.4810$ \\ \hline
$[0/7;9]$ & $-1.0035955$ & $3.6 \times 10^{-3} $ & $ 2.5718$ \\ \hline
$[0/8;7]$ & $-0.9951663$ & $4.8 \times 10^{-3} $ & $ 2.3996$ \\ \hline
$[0/9;7]$ & $-1.0015320$ & $1.5 \times 10^{-3} $ & $ 2.5192$ \\ \hline
$[1/6;7]$ & $-0.9952671$ & $4.7 \times 10^{-3} $ & $ 2.4076$ \\ \hline
$[1/6;8]$ & $-1.0013204$ & $1.3 \times 10^{-3} $ & $ 2.5237$ \\ \hline
$[1/7;6]$ & $-0.9924817$ & $7.5 \times 10^{-3} $ & $ 2.3586$ \\ \hline
$[1/8;6]$ & $-0.9987880$ & $1.2 \times 10^{-3} $ & $ 2.4668$ \\ \hline
$[2/5;7]$ & $-0.9925676$ & $7.4 \times 10^{-3} $ & $ 2.3391$ \\ \hline
$[2/6;6]$ & $-0.9958472$ & $4.2 \times 10^{-3} $ & $ 2.4235$ \\ \hline
$[2/6;7]$ & $-0.9958755$ & $4.1 \times 10^{-3} $ & $ 2.4243$ \\ \hline
$[2/6;8]$ & $-1.0031308$ & $3.1 \times 10^{-3} $ & $ 2.5572$ \\ \hline
$[2/7;5]$ & $-1.0018189$ & $1.8 \times 10^{-3} $ & $ 2.5643$ \\ \hline
$[2/7;6]$ & $-0.9959874$ & $4.0 \times 10^{-3} $ & $ 2.4265$ \\ \hline
$[2/8;6]$ & $-1.0000820$ & $8.2 \times 10^{-5} $ & $ 2.4837$ \\ \hline
$[3/6;4]$ & $-1.0099574$ & $1.0 \times 10^{-2} $ & $ 2.7256$ \\ \hline
$[3/6;5]$ & $-1.0042365$ & $4.2 \times 10^{-3} $ & $ 2.6064$ \\ \hline
$[3/6;6]$ & $-0.9960860$ & $3.9 \times 10^{-3} $ & $ 2.4283$ \\ \hline
$[3/6;7]$ & $-1.0029001$ & $2.9 \times 10^{-3} $ & $ 2.5645$ \\ \hline
$[3/7;6]$ & $-0.9998642$ & $1.4 \times 10^{-4} $ & $ 2.4976$ \\ \hline
$[4/6;4]$ & $-1.0036469$ & $3.6 \times 10^{-3} $ & $ 2.6153$ \\ \hline
$[4/6;5]$ & $-1.0083269$ & $8.3 \times 10^{-3} $ & $ 2.5168$ \\ \hline
$[4/6;6]$ & $-0.9983598$ & $1.6 \times 10^{-3} $ & $ 2.4518$ \\ \hline
$[5/4;6]$ & $-1.0096131$ & $9.6 \times 10^{-3} $ & $ 2.4347$ \\ \hline
$[5/5;6]$ & $-1.0004620$ & $4.6 \times 10^{-4} $ & $ 2.4251$ \\ \hline
$[5/6;4]$ & $-0.9932140$ & $6.8 \times 10^{-3} $ & $ 2.4105$ \\ \hline
$[5/6;5]$ & $-0.9975739$ & $2.4 \times 10^{-3} $ & $ 2.4589$ \\ \hline
\end{tabular}
\end{center}
\caption{Values of $y_{sing}$ and $\gamma_{\ell,a}'$ from differential
approximants to low-temperature series for $\bar\chi^{(a)}_r(y)$. See text 
for definition of $[L/M_0,M_1]$ approximant.  We only display entries which
satisfy the accuracy criterion $|y_{sing}-y_{\ell}| \le 10^{-2}$, where
$y_{\ell}=-1$.}
\label{chiatable}
\end{table}

\subsection{Comment on Exponent Relations at $z=-1$}

   Using the exact results $\alpha_{\ell,FM}'=2$ from eq. (\ref{alphaellp}) 
and $\beta_{\ell}=-1/4$ from eq. (\ref{betal}), and our
conclusion from series analysis that $\gamma_{\ell,FM}'=5/2$, we find that 
\beq
\alpha_{\ell,FM}' + 2\beta_{\ell} + \gamma_{\ell,FM}' = 4
\label{abg}
\eeq
The right-hand side of eq. (\ref{abg}) is twice the value at physical critical
points.  We have given an explanation above of why the exponents
$\alpha_{\ell,FM}'$ and $\beta_{\ell}$ for the singularity at $z=-1$ on the
honeycomb lattice have twice the values of the respective exponents on the
triangular lattice, at the point $u=-1/3$ which corresponds, via
eq. (\ref{uzp}) to $z'=z=-1$ on the honeycomb lattice; this followed from the
star-triangle relation connecting the Ising model on these two lattices
together with the fact that the Taylor series expansion of $u+1/3$, as a
function of $z'$, starts at quadratic order.  We have also noted above the
connection of our finding from the series analysis that $\gamma_{\ell}'$ 
for the honeycomb lattice has twice the value of the corresponding 
$\gamma'$ exponent for the singularity at $u=-1/3$ on the triangular
lattice with the exact relation (\ref{fisherlow}). 
Since each of the exponents on the left-hand side of eq. (\ref{abg})
has twice the value of the respective exponent for the corresponding
singularity at $u=-1/3$ on the triangular lattice, and since this is a linear
equation, the right-hand side is also twice the value of 2 which holds for the
triangular lattice. 

   One may also consider the analogous equation for the approach to $z=-1$ from
within the complex-temperature extension of the AFM phase.  We have extracted
the exact value $\alpha_{\ell,AFM}'=2$ in eq. (\ref{alphaellp}) and, as
discussed above, given the loose-packed nature of the honeycomb lattice and the
resultant relation (\ref{mstag}), it follows that the staggered 
magnetisation diverges with the exponent 
$\beta_{\ell,st}=\beta_{\ell}=-1/4$ as one 
approaches the point $z=-1$ from within the complex-temperature AFM phase.
Combining these exact results with the conclusion from our analysis of the
low-temperature series for the staggered susceptibility that 
$\alpha_{\ell,a}'=5/2$, we find
\beq
\alpha_{\ell,AFM}'+2\beta_{\ell,st}+\gamma_{\ell,a,}'=4
\label{abgstag}
\eeq
in complete analogy with eq. (\ref{abg}), as expected for a loose-packed
lattice.

\subsection{Critical Amplitude for $\chi^{(a)}$ at $z=-1$}
\label{ampchia}

   In order to calculate the critical amplitude $A_{\ell,a}'$ in the 
staggered susceptibility as one approaches $z=y=-1$ from the complex AFM phase,
we compute the series for $(\bar\chi^{(a)}_r)^{1/\gamma_{\ell,a}'}$ using our 
inferred value $\gamma_{\ell,a}'=5/2$.  Since the exact function
$(\bar\chi^{(a)}_r)^{1/\gamma_{\ell,a}'}$ has a simple pole at $z=-1$, one 
performs the Pad\'{e} analysis on the series itself instead of its logarithmic
derivative. The residue at this pole is 
$-y_{\ell}(A_{\ell,a,r}')^{1/\gamma_{\ell,a}'}$,
where $A_{\ell,a,r}'$ denotes the critical amplitude for $\bar\chi^{(a)}_r$.  
Extracting $A_{\ell,a,r}'$ and multiplying by the prefactor, we obtain 
$A_{\ell,a}'=4y_{\ell}^3A_{\ell,a,r}'=-4A_{\ell,a,r}$: 
\beq
A_{\ell,a}' = -0.700 \pm 0.010
\label{ampella}
\eeq

   An analysis of the series for $\bar\chi_r^{1/\gamma_{\ell}'}$ to extract the
critical amplitude for the (uniform) susceptibility as one approaches $z=-1$
from within the complex FM phase did not yield precise results, presumably
because of the shortness of the series.  However, having already inferred that
$\bar\chi$ and $\bar\chi^{(a)}$ have the same power-law divergence at $z=y=-1$,
as approached from the complex FM and AFM phases, respectively, we can use the
relation (\ref{fisherlow}) to compute $A_{\ell}'$ indirectly.  For this
purpose, we recall that on the triangular lattice, at the corresponding 
point $u=u_e=-1/3$, the (uniform) susceptibility $\bar\chi_t$ has the
leading singularity \cite{g75,chitri}
\beq
  \bar\chi_t \sim A_{e,t}'(1+3u)^{-5/4}
\label{chit}
\eeq
Using this, together with the Taylor series expansion (\ref{uzpm1}), we find
the following relations among the critical amplitude $A_{e,t}'$ at $u=-1/3$ 
on the triangular lattice and $A_{\ell}'$ and $A_{\ell,a}'$ at $z=-1$ 
on the honeycomb lattice: 
\beq
  2 \ 3^{5/4} A_{e,t}' = A_{\ell}' + A_{\ell,a}'
\label{amprel}
\eeq
We next use our previous determination of $A_{e,t}'$ \cite{chitri}
\beq
A_{e,t}' = -0.05766 \pm 0.00015
\label{aenum}
\eeq
which agrees with, and has somewhat smaller uncertainty than, an earlier 
determination of this quantity by Guttmann \cite{g75}.  Substituting
(\ref{ampella}) and (\ref{aenum}) into (\ref{fisherlow}), we obtain the
critical amplitude for the uniform susceptibility, 
\beq
 A_{\ell}' = 0.245 \pm 0.010
\label{ampell}
\eeq 

\section{Other Singularities}
\label{zs}

    We also used the low-temperature series for $\bar\chi$ and $\bar\chi^{(a)}$
to investigate the singular behaviour of these functions as one approaches the
points $z = z_{s\pm} = \pm i$ from within the complex-temperature FM and AFM 
phases, respectively.  However, we were not able to obtain conclusive
results.  This is similar to our previous experience investigating the
behaviour of $\bar\chi$ on the triangular lattice in the vicinity of the
singular point $u=-1$ as approached from within the complex FM phase.  These
points, i.e., $z=\pm i$ on the honeycomb lattice and $u=-1$ or equivalently 
$z=\pm i$ on the triangular lattice, share in common the property that they are
intersection points where two arcs of the complex-temperature phase boundary
curves cross.

    Concerning singularities of $\bar\chi$ and $\bar\chi^{(a)}$ elsewhere on
the boundaries of the complex-temperature FM and AFM phases, respectively, it
is quite possible that these functions may exhibit discontinuities, as we found
for $C$ and $M$.  As is well-known from work on first-order phase transitions,
the analysis of the low-temperature series by itself, and similarly, the 
high-temperature series by itself, does not, in general, provide a sensitive 
probe for such discontinuities.  
Of course if one had sufficiently long series, the comparison
of the limits as one approached a phase boundary from, say, the FM and the PM
phases using the respective low- and high-temperature series, could be of value
in this regard.

\section{Behaviour of $\bar\chi$ in the Symmetric Phase }
\label{pmsection}

   The theorem proved in Ref. \cite{ms} and discussed further in
Ref. \cite{chisq} implies that, for the Ising model on the 
square lattice, $\bar\chi$ has at most
finite non-analyticities as one approaches the boundary of the
complex-temperature extension of the PM phase, aside from the physical critical
point at $v=v_c$.  We would expect a similar 
theorem to hold for the honeycomb lattice, although to show this with 
complete rigour, it would be desirable to perform an analysis of the 
asymptotic behaviour of the general connected 2-spin correlation function 
$<\sigma_{0,0}\sigma_{m,n}>$ as $r=(m^2+n^2)^{1/2} \to \infty$ for this
lattice, which has not, to our knowledge, been done.  Assuming that such a
theorem does hold, it would follow, in particular, that, $\bar\chi(v)$ and 
$\bar\chi^{(a)}(v) = \bar\chi(-v)$ would have finite non-analyticities as one
approaches the points $v = \pm i$ from within the PM phase.  We have analysed
the high-temperature series expansion for $\bar\chi(v)$ 
to investigate the singularities at $v = \pm i$.  This series is of the form 
\beq
\bar\chi = 1+\sum_{n=1}^{\infty}a_n v^n
\label{chivseries}
\eeq
It has a finite radius of convergence and, by analytic continuation from the
physical high-temperature interval $0 \le v < v_c$, applies throughout 
the complex extension of the PM phase.  The high-temperature series expansion
(\ref{chivseries}) is known to $O(v^{32})$ \cite{thigh,gaunt}.  Since we
anticipated a finite singularity, we analysed this series using differential
approximants, which are capable of representing this type of singularity in 
the presence of an analytic background term.  However, as we recall from
Fig. 1(c), there are two semi-infinite line segments which protrude into the 
complex-temperature PM phase, with endpoints at $v=\pm v_e = \pm i/\sqrt{3}$. 
We found that the series are not sensitive to the singularities at $v = \pm i$,
presumably because of the effect of the intervening singular line segments and
their endpoints at $v = \pm i\sqrt{3}$. 
We have also tried to study the singularities in $\bar\chi$ at these
endpoints. Again, our study did not yield 
an accurate value for the exponent $\gamma_e$, presumably due to the 
insufficient length of the series.  However, the (scattered) values of
$\gamma_e$ were consistent with the expectation that $\gamma_e < 0$.

\section{Conclusions} 

   In this paper we have investigated complex-temperature singularities in the
Ising model on the honeycomb lattice.  As part of this, we have discussed the 
complex-temperature phases and their boundaries.  From exact results, we have
determined these singularities completely for the specific heat and the 
uniform and staggered magnetisation.  
From an analysis of low-temperature series expansions, we have found 
evidence that $\chi$ and $\chi^{(a)}$ both have divergent singularities at 
$z=-1 \equiv z_{\ell}$ (where $z=e^{-2K}$), with exponents 
$\gamma_{\ell}'= \gamma_{\ell,a}'=5/2$. The critical amplitudes at this 
singularity were calculated.  We have found that the linear combination of
exponents $\alpha'+2\beta+\gamma'$ is equal to 4 rather than 2 at $z=-1$.  The
connection of these results to corresponding complex-temperature singularities
on the triangular lattice was discussed.  Finally, some results on
complex-temperature singularities reached from within the symmetric phase were
given. 

\vspace{4mm}

   One of us (RS) would like to thank Profs. David Gaunt and Tony Guttmann for
information about the current status of the series expansions for the honeycomb
lattice.  This research was supported in part by the NSF grant PHY-93-09888.

\vfill
\eject

\begin{center}
{\bf Figure Caption}
\end{center}

\vspace{5mm}

Fig. 1. \  Complex-temperature phases and associated boundaries in the 
variables (a) $z$, (b) $k_<$, and (c) $v$.  FM, AFM, and PM denote 
ferromagnetic, antiferromagnetic, and paramagnetic, $Z_2$-symmetric phases. In
Fig. 1(b), the semi-infinite line segment extends along the negative real axis
from $k_<=0$ to $k_< = -\infty$.  In Fig. 1(c), the semi-infinite line segments
extend along the imaginary axis from $v=\pm i/\sqrt{3}$ to $\pm i\infty$, 
respectively. 
\vfill
\eject

\end{document}